\documentclass[manuscript,nonacm]{acmart} %review nonacm manuscript

\usepackage{soulutf8}
\usepackage{subcaption}
\usepackage{multirow}

\begin{document}

\title{Weighted Multi-Level Feature Factorization for App ads CTR and installation prediction }

\author{Juan Manuel Rodriguez}
\email{juanmanuel.rodriguez@isistan.unicen.edu.ar}
\affiliation{%
%\institution{\small{juanmanuel.rodriguez@isistan.unicen.edu.ar}}
  \institution{Aarhus University}
  %\city{City}
  \country{Denmark}
}\affiliation{%
%\institution{\small{juanmanuel.rodriguez@isistan.unicen.edu.ar}}
  \institution{ISISTAN, CONICET-UNCPBA}
  %\city{City}
  \country{Argentina}
}

\author{Antonela Tommasel}
\email{antonela.tommasel@isistan.unicen.edu.ar}
\affiliation{%
%\institution{\small{antonela.tommasel@isistan.unicen.edu.ar}}
  \institution{ISISTAN, CONICET-UNCPBA}
  % \city{City}
  \country{Argentina}
}

\begin{abstract}
This paper provides an overview of the approach we used as team ISISTANITOS for the ACM RecSys Challenge 2023. 
The competition was organized by ShareChat, and involved predicting the probability of a user clicking an app ad and/or installing an app, to improve deep funnel optimization and a special focus on user privacy.
Our proposed method inferring the probabilities of clicking and installing as two different, but related tasks. 
Hence, the model engineers a specific set of features for each task and a set of shared features. 
%This engineering uses a series of linear layers consider different orders of interaction of the features.    
Our model is called Weighted Multi-Level Feature Factorization because it considers the interaction of different order features, where the order is associated to the depth in a neural network. The prediction for a given task is generated by combining the task specific and shared features on the different levels.  
% By combining the task specific features with the shared features, the model generates the prediction for the given task.
Our submission achieved the 11 rank and overall score of 55 in the competition academia-track final results.
We release our source code at: \url{https://github.com/knife982000/RecSys2023Challenge}

\end{abstract}

\maketitle

\section{Introduction}

Recommender systems have emerged as influential tools across various platforms, aiding users in discovering interesting items and ultimately influencing their consumption patterns. %They aim at predicting the preference of users for specific elements, based, mostly, on their previous interests and behaviour. 
Since the early 2000s, online advertising has developed into a thriving industry worth billions of dollars, playing a significant role in Internet’s growth and offering a common marketing experience to people accessing online services \cite{gharibshah2021user}. The main difference between online and traditional mass advertising is its inherent ability to tailor messages to individual users, democratizing advertising and enabling businesses of all sizes to participate. %Moreover, it provides advertisers with the ability to measure the impact of their expenditure. 
Accurately predicting clicks, as the first measurable user response, is a crucial step for many digital advertising and recommenders to capture the user likelihood of users taking subsequent actions, such as purchasing a product, subscribing to a service, or, as in this case, installing an app \cite{gharibshah2021user}.

The ACM Recsys Challenge 2023\footnote{\url{http://www.recsyschallenge.com/2023/}} was organized by ShareChat, and involved predicting the probability of a user installing an app, intending to improve deep funnel optimization and a special focus on user privacy. To this end, we proposed a model that generates prediction using different order features~\cite{wdlrs16, deepfm17} and 
 a weighted linear combination, and it is called Weighted Multi-Level Feature Factorization.

% This paper is organized as follows. 
% Section~\ref{sec:problem} presents the problem definition and describes the data collection and evaluation metrics. 
% Section~\ref{sec:data-eng} describes the data engineering process. 
% Then, Section~\ref{sec:model} details the proposed model for predicting app installation likelihood. 
% Section~\ref{sec:evaluation} analyzes the results of the performed experimental evaluation.
% To explore the generalizability of the proposed model, Section~\ref{sec:evaluation-other} reports additional experimental evaluation carried out over other data collections.
% Finally, Section~\ref{sec:conclusions} presents the conclusions and future works.

% Challenge Task top
% Online advertising has been a multi-billion dollar industry since the early 2000 and has played a significant role in the growth of the internet. The key advantage of online advertising over conventional mass advertising is its inherent ability to personalize to users, democratizing advertising and enabling businesses of all sizes to participate, and providing the measurable impact of money spent to the advertisers. Over the past two decades, the nature of online advertising has also evolved tremendously from pure banner-based advertising, where advertisers were charged based on the number of ad impressions, to deep funnel optimizations, where advertisers can optimize for eventual sales.

\section{Problem formulation}\label{sec:problem}

The ACM RecSys Challenge 2023, organized by ShareChat\footnote{\url{ https://sharechat.com/about}} aims to predict whether a user would install an advertised app on the next day, given user and ad features and data from impressions, clicks and installs from the past 2 weeks. 
%hat was advertised to them, based on user and ad features, and user-ad interaction data. The goal is, given data from impressions, clicks and installs from the past 2 weeks, predict the likelihood of a user installing the advertised app on the 15th day. 
The task is proposed in the context of deep funnel optimization with a special interest in preserving users’ privacy. 

\subsubsection*{\textbf{Data collection.}}\hfill

The data collection includes information for approximately 10 million random users who visited the ShareChat and Moj apps for over 3 months. For each user, there are 10 ad impressions available. The collection includes features user, ads, and user-ad interaction features. \textit{User features.} include i) demographic features, like age, gender, and anonymized user location. User sampling guarantees an approximate uniform distribution of these demographic features. ii) content preference embeddings trained based on users’ consumption of the non-ad content of the ShareChat and Moj apps. iii) embeddings trained based on the past apps installed by the user through the platform.
\textit{Ad features} include i) categorical anonymized features representing, for example size and category. ii) emdeddings representing ad's  actual content (video/image). Finally, \textit{historical interactions} include count features representing the user interaction with ads, advertisers and categories of advertisers over different lengths of a time window

% \noindent\textbullet~{\textit{User features.}} These include i) demographic features, such as age, gender, and anonymized user location. User sampling guarantees an approximate uniform distribution of these demographic features. ii) content preference embeddings trained based on users’ consumption of the non-ad content of the ShareChat and Moj apps. iii) app affinity embeddings trained based on the past apps installed by the user through the platform.

% \noindent\textbullet~{\textit{Ad features.}} These include i) categorical anonymized features representing, for example size and category. ii) emdeddings representing the actual content of the ad (video/image).

% \noindent\textbullet~{\textit{Historical interactions.}} These include count features representing the user interaction with ads, advertisers and categories of advertisers over different lengths of a time window. %Every instance has an associated numeric id that represents an ad impression shown to the user and whether it resulted in a clicked ad and an installed app.

\smallskip
Then, each data instance is composed by 82 features: $f_{0}$ instance id, $f_{1}$ date of the impression, $f_{2}$ to $f_{32}$ categorical features, $f_{33}$ to $f_{41}$ binary features, $f_{42}$ to $f_{79}$ numeric features, and $\text{is\_clicked}$ and $\text{is\_installed}$ binary labels. The challenge presents two different, yet related tasks, inferring the $\text{is\_clicked}$ and $\text{is\_installed}$ labels. %The relationship between the task can be appreciated by analizing the labels on the training dataset. 
Regarding class distribution, 21.98\% instances were clicked, while 17.40\% where installed and only 7.11\% were both clicked and installed.
% From all the instances, 21.98\% were clicked and 17.40\% where installed, but when analyzing both labels at the same time it can be seen that 7.11\% were both clicked and then installed, i.e., 32.36\% of the instances clicked where installed, which is higher than the installed  proportion for the whole dataset. 
These tasks, complex by themselves, present some additional challenges: i) the lack of semantic information of the individual features; ii) the limited number of interactions known for each user; 
and iii) there is no distinction of which features represent the user and which ones the ad. %As well as in previous editions, great importance has been given to data leakage. Using data from validation or test partitions to compute features or train the models is not allowed. In addition, using external data or models trained with external data is not allowed either.

\subsubsection*{\textbf{Evaluation goal.}}\hfill

%Given the purpose of predicting whether users will install an app based on an ad impression, 
Predictions are used to generate rankings that measure the expected revenue for the platform and then the winner ad is displayed to the user. %Since predictions are used in the actual ranking, it is important for probabilities to be accurate. 
Recommendations were evaluated by means of normalized Cross Entropy.

\section{Data engineering}\label{sec:data-eng}
Different pre-processing strategies were applied to the features based on their type. %categorical or numerical. Binary features were treated as categorical. 
After processing, each instance was represented with two feature vectors, one for the categorical ($X_c$) and one for the numeric ($X_n$) features. Impression date ($f_{1}$) was ignored as its value is constant in the test set and such value is not present in the training set.

\subsubsection*{\textbf{Categorical features}}\hfill 

Two types of transformations were applied based on the number of categories. For features with over 20 categories, an ordinal encoding process was followed, assigning items IDs from 0 to $\#items -1$ to facilitate the feature processing by the neural network. Null and categories not existing in the training set were assigned 
%If in the test set appeared a category that was not present on the training set or that had a null value, it was assigned 
the ID of the most frequent category. Although this strategy might affect the relation of the most frequent categories with the others, less than 0.74\% were affected in the worst case, which might not be enough to cause any negative effect during training or inference.

Categorical features with fewer than 20 categories were transformed into two numeric features, which were defined as the probability of being clicked or being installed given the value of the feature. Probabilities were estimated by computing the percentage of ad clicks and app installs in the training dataset for each value. For example, Eq.~\ref{eq:TC} and Eq.~\ref{eq:TI} show the transformation for $f_{32}$. Similar to the previous case, categories in the test set that were not present in the training were replaced by the most frequent category.\vspace{-0.75cm}

% \begin{equation}\label{eq:TC}
%     TC_{f_{32}} \left(c\right) = P\left(\text{is\_clicked } | f_{32} = c\right)
% \end{equation}

% \begin{equation}\label{eq:TI}
%     TI_{f_{32}} \left(c\right) = P\left(\text{is\_installed } | f_{32} = c\right)
% \end{equation}

\begin{table}[H]
    \centering
   
\begin{tabular}{cc}

$ TC_{f_{32}} \left(c\right) = P\left(\text{is\_clicked } | f_{32} = c\right) \hspace{0.3cm}\refstepcounter{equation}(\theequation)\label{eq:TC}$ 
& 
$  TI_{f_{32}} \left(c\right) = P\left(\text{is\_installed } | f_{32} = c\right)
\hspace{0.3cm}\refstepcounter{equation}(\theequation)\label{eq:TI}$

    \end{tabular}\vspace{-0.5cm}
   
\end{table}

\subsubsection*{\textbf{Numerical features}}\hfill 

Missing numeric features ($f_42$ to $f_79$) were first imputed with their average value. Then, they were standardized according to $ S\left(c\right) = \frac{c-\mu}{\lambda\sigma}$, where $\mu$ is the average feature value, $\sigma$ is the standard deviation and $\lambda$ is a hyperparameter set to 3. This value was chosen so, in the case of a normal distribution, $99.9\%$ of values would be at most $3\sigma$ of $\mu$, i.e., those values would be between -1 and 1.%\vspace{-0.5cm}

\begin{figure*}
    \centering

    \subfloat[Original feature distribution\label{fig:original_f60}]{\includegraphics[width=0.30\textwidth]{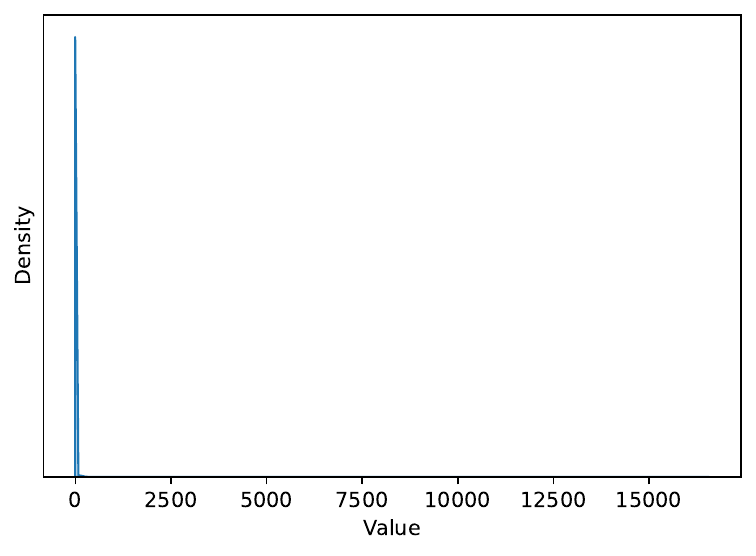}}~
    \subfloat[Transformed feature distribution\label{fig:transformed_f60}]{\includegraphics[width=0.30\textwidth]{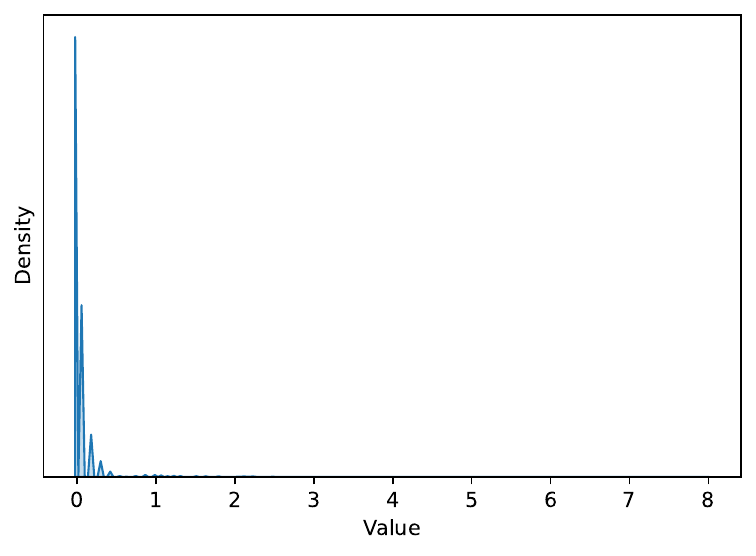}}\vspace{-0.2cm}

    \caption{Kernel Density Estimation of feature $f_60$ before and after transformation}
    \label{fig:feature-transformation}\vspace{-0.5cm}
\end{figure*}

Considering that most features contained extreme values and were also highly sparse (for $31\%$ of numeric features over $90\%$ of the values were zero), only applying the first transformation caused features to have extreme values that might be difficult to process by ML models. For example, while feature $f_{60}$ contained $98.36\%$ zeros in the training set with a mean of 0.889 ($\pm$ 43.27) and $9.91\%$ lower than $\mu + 3\sigma$, its highest value was 16,570.97. To handle this case, we transformed the features as Eq. \ref{eq:L2} shows (where $t$ represents $S(c)$) to obtain a continuous function that it is linear when $t$ is in the range $[-1,1]$ and logarithmic outside of that range. As an example, Figure \ref{fig:feature-transformation} shows the changes in minimum and maximum values of $f_60$ after applying the described transformations.\vspace{-0.3cm}

% \begin{equation}\label{eq:L1}
% L\left(t\right) = \begin{cases}
%                         t & abs\left(t\right) < 1\\
%                         sign\left(t\right)log_2\left(abs\left(t+1\right)\right) & abs\left(t\right) \geq 1\\
% \end{cases}
% \end{equation}

\begin{equation}\label{eq:L2}
L\left(t\right) = \begin{cases}
                        t & -1 < t < 1\\
                        log_2\left(t+1\right) & t \geq 1\\
                        -log_2\left(-t+1\right) & t \leq -1\\
\end{cases}
\end{equation}\vspace{-0.95cm}

\section{Method}\label{sec:model}

\begin{figure*}
     \centering
    \includegraphics[width=0.98\textwidth]{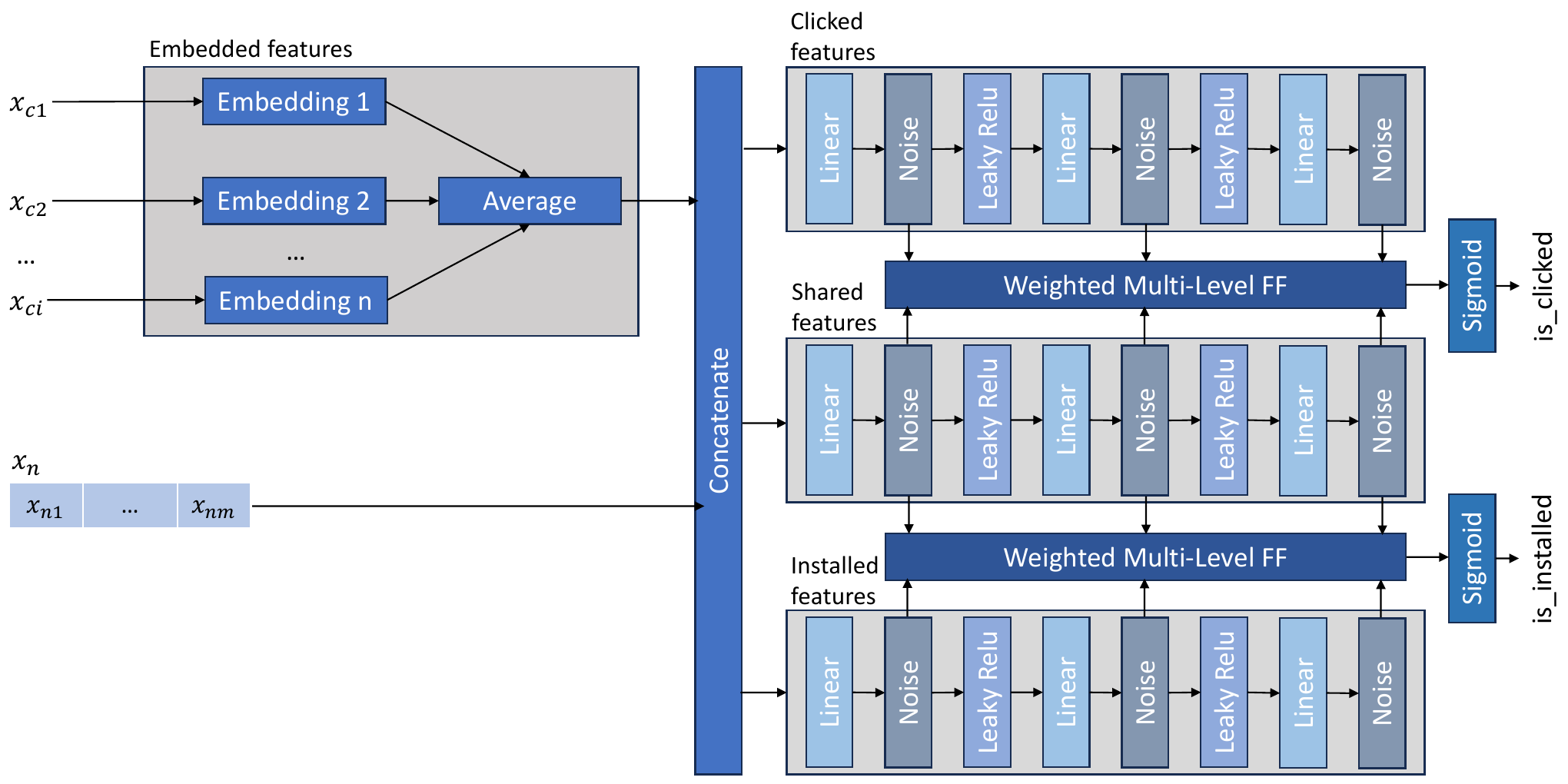}\vspace{-0.2cm}
    \caption{Schematic diagram of the proposed model}
    \label{fig:model}\vspace{-0.54cm}
\end{figure*}

The overall architecture of the proposed model is schematized in Figure~\ref{fig:model}. Our model uses an operation similar to matrix factorization to compute feature interactions. Instead of directly estimating the matrices involved, sequences of linear layers compute different order features interactions by estimating one feature vector per linear layer. Then, several predictions are generated by computing the dot product between different level vectors and then weighed and added to generate a final prediction. This solves the drawback of matrix factorization requiring specific data input, making it unsuitable for general purpose prediction tasks~\cite{fm2010}. The model takes inspiration on Wide and Deep~\cite{wdlrs16} and DeepFM~\cite{deepfm17} architectures, which recognized the importance of both capturing low-order interaction between features (i.e., Wide or Factorization Machines~(FM) part) and learning high-order interactions (i.e., Deep learning part).

The model is divided into 4 components: \textit{embedded}, \textit{click}, \textit{shared} and \textit{installed} \textit{features}. \textit{Embedded features} is responsible for transforming the categorical categories into a dense representation. After transformations, each categorical feature is embedded, and the resulting embeddings are averaged to obtain the final categorical feature vector. Finally, the obtained vector is concatenated with the numeric features vector to obtain the dense instance representation. This vector is fed to the \textit{click}, \textit{shared} and \textit{installed Features} components. These components share the same architecture and are defined as a sequence of dense and noise layers and an activation function. The noise layer is only applied during the training phase and is defined as Eq.~\ref{eq:noise} where $\mathcal{N}(1,\sigma_\lambda)$ represents a sampling over a normal distribution with $\mu = 1$, and a standard deviation $\sigma_\lambda = 0.5$. This noise ensures that the mean absolute percentage error (MAPE) between the feature and the noise feature remains constant independently of the scale of the feature. Moreover, it acts as a regularizer, as the noise values affecting features with large absolute values are larger than those for features with small absolute values. Based on the dense instance representation, these components aim at generating the most relevant features for the prediction. Each component returns 3 vectors, one for each linear layer, representing features at different depths, namely $F^{t}=(f^{t}_{1}, f^{t}_{2}, f^{t}_{3})$ where $t\in{c, s, i}$ is the type of feature and the sub-index is the layer depth that generated the feature. Using \textit{click features} and \textit{shared features}, our model infers the probability of $is\_clicked$; while \textit{installed features} and \textit{shared features} are used to infer the probability of $is\_installed$. The goal of the \textit{shared features} component serves to represent the similarities between both tasks, and to limit the number of parameters, reducing the variance of the model.\vspace{-0.3cm} 

% \begin{equation}\label{eq:noise}
%     noise\left(x\right) = \begin{cases}
%                             x + \mathcal{N}\left(0,\sigma_\lambda\right)& \text{train}\\
%                             x & \text{inference}
%                           \end{cases}
% \end{equation}
\begin{equation}\label{eq:noise}
    noise\left(x\right) = x\mathcal{N}\left(1,\sigma_\lambda\right) %Lo cambien porque es sabido que noise solo se aplica en training
\end{equation}

% \hl{[FALTA MUCHO DETALLE ACÁ]} The goal of having two specific and one shared layer \hl{[QUEDA MUY DESCOLGADO, NO QUEDA CLARO A QUÉ TE ESTÁS REFIRIENDO]} is two-fold. First, given that there is no separation between the user and item features, these layers allow learning the feature interactions from the combined set of features. Second, they allow reducing variance by reducing the number of parameters and \hl{[ y relaciona internamente las dos tares a realizar – NO ESTÁN EXPLICADAS LAS DOS TAREAS: lo agregue en la descripción del dataset. Las dos tareas son predecir cada uno de los labels]}
After generating the features, the model combines the feature vectors of each level using a dot product (similar to a matrix factorization operation), weights predictions with a learned parameter, and scales predictions using a shared parameter for all predictions. Our model defines the \textit{Weighted Multi Level Feature Factorization}~(WMLFF) ad-hoc layer to perform this operation. WMLFF is defined in Eq.~\ref{eq:WMLFF}, where $m$ is a scalar parameter shared by all WMLFF layers and $w_l$ is a learned weight parameter for that level and the WMLFF layer. Finally, the output of these layers is passed through a sigmoid function to estimate the clicked/installed probability. In brief, Eq.~\ref{eq:is_click} defines the inference for $is\_clicked$ label, and Eq.~\ref{eq:is_installed} defines the inference for $is\_installed$ label.\vspace{-0.3cm}
%matrix factorization operations with the features at different depth from the Click features, Shared features and Installed features \hl{[COMO DE LAS CAPAS? NO LOS LLAMABAS COMPONENTES AL PRINCIPIO?]}. In this layers, $L$ represents the number of levels, $p_l$ and $s_l$ represent the specific and shared features in level $l$, $\langle p_l,s_l \rangle$  is the internal product of features, $w_l$ is a weight parameter during learning and m is a multiplier shared by all WMFL that is also leaning during training. Finally, the outputs of the WMLF layers are passed through a sigmoid function to estimate the probability of an ad being clicked and an app being installed.

% \begin{equation}\label{eq:WMLFF}
%     WMLFF\left( F^{t1}, F^{t2} \right) = m \sum_{l=1}^{L} w_l \langle f^{t1}_l,f^{t2}_l \rangle
% \end{equation}
% \begin{equation}\label{eq:is_click}
%     \hat{is\_clicked} = \sigma\left( WMLFF\left( F^{c}, F^{s} \right)\right)
% \end{equation}
% \begin{equation}\label{eq:is_installed}
%     \hat{is\_installed} = \sigma\left( WMLFF\left( F^{i}, F^{s} \right)\right)
% \end{equation}

\begin{table}[H]
    \centering
   
\begin{tabular}{cc}

\multirow{2}{0.5\columnwidth}{ $WMLFF\left( F^{t1}, F^{t2} \right) = m \sum_{l=1}^{L} w_l \langle f^{t1}_l,f^{t2}_l \rangle \hspace{0.3cm}\refstepcounter{equation}(\theequation)\label{eq:WMLFF}$  } 

& $  \hat{is\_clicked} = \sigma\left( WMLFF\left( F^{c}, F^{s} \right)\right)\hspace{0.3cm}\refstepcounter{equation}(\theequation)\label{eq:is_click}$ \tabularnewline
& $   \hat{is\_installed} = \sigma\left( WMLFF\left( F^{i}, F^{s} \right)\right)
\hspace{0.3cm}\refstepcounter{equation}(\theequation)\label{eq:is_installed}$

    \end{tabular}\vspace{-0.5cm}
   
\end{table}

\subsubsection*{\textbf{Model training}}\hfill 

A binary cross-entropy loss function (Eq.~\ref{eq:loss}) reduced by mean is used, where $N$ is the number of predictions in the batch, $c$ and $\hat{c}$ represent the ground truth and predictions for $is\_clicked$ and $i$ and $\hat{i}$ represent the ground truth for $is\_installed$.\vspace{-0.3cm}

\begin{equation}\label{eq:loss}
    L\left(c,\hat{c},i,\hat{i}\right) = \frac{
                -\sum c \cdot log\left( \hat{c} (\right)
                + \left(1-c\right) \cdot log\left(1-\hat{c}\right)
                + i \cdot log\left( \hat{i} \right)
                +\left(1-i\right) \cdot log\left(1-\hat{i}\right)
                }{2N} 
\end{equation}

% L(c,c ̂,i,i ̂ )=(-∑c⋅log(c ̂ )+(1-c)⋅log(1-c ̂ )+i⋅log(i ̂ )+(1-i)⋅log(1-i ̂))/2N

\subsubsection*{\textbf{Implementation details}}\hfill 

The model was implemented in PyTorch \footnote{\url{https://pytorch.org}}. The optimizer was RAdam\footnote{\url{https://pytorch.org/docs/stable/generated/torch.optim.RAdam.html}} using the default configuration. All embedding and linear layers output sizes were set to 32. Batch size was set to 1024. The learning process ran for 40 epochs.

\section{Experimental evaluation}\label{sec:evaluation}

Table~\ref{tab:results} presents the results obtained for the proposed model. These results are 5.93\% lower than the best submission for the best submission of the academic leaderboard. Aiming at improving the obtained results, we also report different variants of the proposed model that alter its architecture or hyperparameters: 

% \begin{itemize}
\noindent\textbullet~\textit{Original with $\sigma_\lambda = 0.3$}. As noise acts as a regularizer, the goal of this layer was to reduce over-fitting, however, as noise increases, optimizing the network might be hindered. This variant reduced the standard deviation of the $\mathcal{N}$ distribution set for the noise layer from $0.5$ to $0.3$. As noise was reduced, the result worsened, which might indicate over-fitting.

\noindent\textbullet~\textit{Original with QHAdam\footnote{\url{https://github.com/jettify/pytorch-optimizer}} \& Original with AdamW}. These variants changed the optimizer function. In both cases, results were similar to those of the original model. As we only have access to the final results, whether differences were statistically significant is unknown. 

\noindent\textbullet~\textit{Original without shared features}. The goal of the \textit{shared features} component was to capture the similarities between the two tasks, thus reducing the number of parameters and variance of the model. Nonetheless, this simplification could negatively affect the results if no actual relation existed between the tasks. Therefore, in this variation, two feature components for the $is_installed$ task and two for $is_clicked$ task were included. Results showed that having separated components decreased performance, indicating that there is a correlation between both tasks.%In this regard, this variant removes the Shared features component and one component per type of feature per task is added, resulting in four feature components: Installed 1 features, Installed 2 features, Clicked 1 features and Clicked 2 features. \hl{[POR QUÉ 1 Y 2?? QUÉ SIGNIFICAN? – QUÉ DIO?]}

\noindent\textbullet~\textit{WMLFM based on cosine similarity}. This variant modified the WMLFM module to use cosine similarity instead of the dot product. This variant assumed that the likelihood of an ad being clicked, or an app being installed is influenced by the angle between the feature vectors regardless of their length, i.e., vectors whose angles are close to zero are related to the positive label, while angles close to 180 are related with negative labels. As the Table shows, this variant reduced the performance of the model, which could be related to, for example, the fact that the introduced noise has a major effect over the cosine similarity than the dot product or that cosine similarity causes gradients to be closer to zero, resulting in a vanishing gradient problem.

\noindent\textbullet~\textit{Original with k-fold}. Divided the training set into 10-folds sub-training/validating sets, then 10 different models were trained until the results not improved on the validating set. Predictions were computed as the average of all models outputs, i.e., these model act as an ensemble model. The idea is to have several models better fitted to the training set, but without over-fitting it. Results show that this approach might over-fitted the training dataset. This could be because there is a difference between the data in the train set and the test set that was not captured by the splitting or that the ensemble technique was unappropriated for the model.

\noindent\textbullet~\textit{Original with deeper feature components}. This variation defined the \textit{click}, \textit{shared} and \textit{installed feature} components with six linear layers instead of three layers. This version tests whether adding deeper feature layers improved results. Increasing the layer did not achieve better results on the training loss, or in the test results. This point out that such a deep model is hard to train, which might be results a phenomenon like banishing gradient.

\noindent\textbullet~\textit{Original with 64 dimensions}. All embedding and linear layers output sizes were set to 64. Increasing the number of dimensions allows the model to fit the training dataset better, but at the risk of over-fitting it. During training, the model got a lower loss, but results show that the model over-fitted training as it was outperformed on the test data.
% \end{itemize}

\begin{table}
    \centering
    \scalebox{0.99}{
    \begin{tabular}{l c c c} 

Model & Competition Results\tabularnewline	
\toprule
Best submission company & 5.744\tabularnewline	
Best submission academia & 6.054\tabularnewline	
Original & 6.413\tabularnewline		
Original with $\sigma_\lambda = 0.3$ & 6.65\tabularnewline		
Original with QHAdam optimizer & 6.414\tabularnewline		
Original with AdamW optimizer & 6.421\tabularnewline		
Original without shared features  & 6.471\tabularnewline		
WMLFM based on cosine similarity & 6.602\tabularnewline		
Original with k-fold & 6.649\tabularnewline		
Original with deeper feature components & 6.489\tabularnewline		
Original with 64 dimensions & 6.494\tabularnewline

    \bottomrule 
    \end{tabular}
    }
    \caption{Results obtained for the evaluated models for the ShareChat dataset.}
    \label{tab:results}\vspace{-1.0cm}
\end{table}

\section{Experimental evaluation with other data collections}\label{sec:evaluation-other}

To assess the generalizability of the proposed model, this Section introduces additional evaluations performed over other data collections from the literature.

\subsection{Criteo}

This evaluation considers the Criteo\_x1 dataset\footnote{\url{https://github.com/openbenchmark/BARS/tree/main/datasets/Criteo\#criteo_x1}}~\cite{bars_ds}, which is preprocessed and divided into training, validation and test as described in~\cite{afn_criteo}. Training, validation and test consists of 33,003,326, 8,250,124, and 4,587,167 instances, respectively. Each instance consists of 26 categorical features, 13 continuous features, and a binary label. The three partitions have roughly the same proportion of positive labels.%, training and validation 25.62\% and 25.60\% in test. 
Presented in the Criteo Display Advertising Challenge\footnote{\url{https://www.kaggle.com/c/criteo-display-ad-challenge}}, the task was defined as predicting the probability of a user clicking on ad on a given Web site. The model was modified to have only two feature sub-networks, the loss function was set to cross-entropy, and early stopping was applied.

\begin{table}
    \centering
    \scalebox{0.99}{
    \begin{tabular}{l c c c} 

Model & Test AUC & Test Cross-Entropy (LogLoss) & Best Epoch\tabularnewline
    \toprule
Original & 0.796 & 0.454 & 5\tabularnewline
Original with 64 dimensions & 0.799 & 0.451 & 5\tabularnewline
Original $\sigma_\lambda = 0.25$ & 0.8 & 0.45 & 5\tabularnewline
Original with 64 dimensions $\sigma_\lambda = 0.25$ & 0.802 & 0.448 & 3\tabularnewline
Original $\sigma_\lambda = 0.10$ & 0.802 & 0.449 & 4\tabularnewline
Original with 64 dimensions $\sigma_\lambda = 0.10$ & 0.803 & 0.448 & 3\tabularnewline
Original No Noise & 0.802 & 0.449 & 4\tabularnewline
Original with 64 dimensions No Noise & 0.804 & 0.447 & 3\tabularnewline
\midrule
DeepFM~\cite{deepfm17} & 0.801 & 0.451 & -\tabularnewline
AFN+~\cite{afn_criteo} & 0.807 & 0.445 & -\tabularnewline
FinalMLP2023~\cite{FinalMLP2023} & 0.815 & - & -\tabularnewline
    \bottomrule 
    \end{tabular}
    }
    \caption{Results obtained for the evaluated models for the Criteo dataset.}
    \label{tab:results-criteo}\vspace{-1.0cm}
\end{table}

Table~\ref{tab:results-criteo} presents the result of the experiment. %For this experiment, we modified the model by removing one of the feature component as the Criteo dataset has only one task. 
The version named original uses the same hyper-parameters as the ones used for the task. For the other versions, the number of internal representation dimensions was set to 64, and the noise level was modified. This table also presents the reported results for DeepFM~\cite{deepfm17}, AFN+~\cite{afn_criteo} and FinalMLP2023~\cite{FinalMLP2023}\footnote{State-of-the-art performance according to \url{https://paperswithcode.com/sota/click-through-rate-prediction-on-criteo}. Retrieved: 11th July 2023}. Our results shows that without much tuning our model performs similarly of models from 2017-2019, but it is outperformed by an State-of-the-Art model by less than 2\%. Moreover, increasing the dimensions and removing the noise improves the results pointing out that these hyperparameters effect is tide to the dataset characteristics. 

\subsection{MovieLens 100k}

MovieLens100k dataset~\cite{movielens} is a movie 1 to 5 rating dataset containing a total of 100\,000 ratings given by 943 users on 1682 movies. Given the dataset characteristics, some modifications to the proposed model were introduced. The goal here was to estimate the rating given by a user to a movie. Then, the loss function was changed to MSE (Mean Squared Error). The target feature was transformed from the range 1-5 to 0-1 by subtracting 1 and divided by 4. Later, to compute RMSE. This transformation was not applied for the linear versions where the sigmoid function at the output of the model was removed, and thus the model can return any real value. This was done to fully treat the problem as a regression problem, as when using the sigmoid activation the model restricted the output to values between zero and one%, while removing the sigmoid the model is capable of output any real value
. Given the size of the data, the batch size was set to 64. The original u1\footnote{\url{https://grouplens.org/datasets/movielens/100k/}} division into 80,000 training and 20,000 test instances was used.

From the original dataset, we considered categorical and binary features. The included categorical features were: user, occupation, age, and movie, which were ordinal encoded. Binary features represented the movie genres and user gender, which seems to be related to movie rankings~\cite{gender_ml}. User gender is represented as two mutually exclusive binary features Male and Female. %Notice that movie genres are not mutually exclusive, e.g, GoldenEye (1995) is Action, Adventure, and Thriller. 
Binary features were considered as numeric features as the first dense layer can learn representations for features with a value of 1. Finally, 4 additional features are added to represent the average and the percent standard deviation of the user and item bias. In the case of the non-linear model the obtained average is transformed using the same transformation as the target feature. Percent standard deviation is computed as $\mu / \sigma$.

\begin{table}
    \centering
    \scalebox{0.99}{
    \begin{tabular}{l c} % cccc

Model & RMSE\tabularnewline
\toprule
Original & 0.936\tabularnewline
Original - Linear & 0.939\tabularnewline
Original $\sigma_\lambda = 0.25$ & 0.963\tabularnewline
Original No Noise & 1.102\tabularnewline
Original + feature user\textsubscript(avg)  + user\textsubscript(\%std) + movie\textsubscript(avg)  + movie\textsubscript(\%std) & 0.928\tabularnewline
Original - Linear + feature user\textsubscript(avg)  + user\textsubscript(\%std) + movie\textsubscript(avg)  + movie\textsubscript(\%std) & 0.928\tabularnewline
\midrule
Factorized EAE~\cite{dmias2018} & 0.920\tabularnewline
GHRS~\cite{GHRS2022} & 0.887\tabularnewline
    \bottomrule 
    \end{tabular}
    }
    \caption{Results obtained for the evaluated models for the MovieLens100k dataset}
    \label{tab:results-movie100k}\vspace{-1.0cm}
\end{table}

Table~\ref{tab:results-movie100k} depicts the results of predicting the rating for movies. %Unlike the Criteo dataset, the task is different as it is a regression task rather than a classification task. In this case, 
We present a variation of the model for regression with a known range of output values (original), and unbounded (linear). %The Table presents also a variation of the model regarding the noise level. 
Results showed that high noise improves the performance supporting the idea that this hyperparameter effect is tide to the dataset. Moreover, enriching features by using statistics of the rating improves the results suggesting that this information is useful, which is a similar idea to how certain categorical features were encode for the challenge. For the sake of comparison, we included the reported results of Factorized EAE~\cite{dmias2018} and GHRS\footnote{State-of-the-Art according to \url{https://paperswithcode.com/sota/collaborative-filtering-on-movielens-100k}. Retrieved 11th July 2023}~\cite{GHRS2022}. Despite not being designed for this task, our model performs similarly to a 2018 model, and less than 5\% lower than the State-of-the-Art model.  

% \subsection{MovieLens 1m}

% \hl{[FALTA DETALLE DE TAMAÑO DEL DATASET]}

% \hl{[TIENE SENTIDO REPORTAR ESTE SI CORRISTE SOLO UN MODELO?: Si mejor lo sacamos]}

% We used the MovieLens1m dataset\footnote{\url{https://paperswithcode.com/sota/collaborative-filtering-on-movielens-1m}} following the same procedure as for MovieLens100k. Dataset was temporally split by user, considering the first 90\% of ratings for training and the remaining 10\% for test.

% \begin{table}
%     \centering
%     \scalebox{0.99}{
%     \begin{tabular}{l c} % cccc

% Model & RMSE\tabularnewline		
% \toprule
% Original + feature user\textsubscript(avg)  + user\textsubscript(\%std) + movie\textsubscript(avg)  + movie\textsubscript(\%std) & 0.89\tabularnewline		
% Small batch + feature user\textsubscript(avg)  + user\textsubscript(\%std) + movie\textsubscript(avg)  + movie\textsubscript(\%std) & 0.873\tabularnewline		

%     \bottomrule 
%     \end{tabular}
%     }
%     \caption{Results obtained for the evaluated models for the MovieLens1m dataset}
%     \label{tab:results-movie1m}%\vspace{-1.0cm}
% \end{table}

\section{Conclusions}\label{sec:conclusions}

This paper presents ISISTANITOS's model for the RecSys Challenge 2023. 
Our model is based on learning sets of multi-level features that capture different order of interactions between the input features.
This is based on the concepts presented in previous works~\cite{wdlrs16, deepfm17}, adapted to the particular tasks of the challenge. Finally, we present a preliminary assessment of the concept of multi-level features applied to other recommender system benchmarks.
These results showed that our model achieves good performance on these benchmarks. Hence, more evaluations are needed to assess whether the multi-level feature concept can obtain State-of-the-Art results in different recommender system benchmarks.

\section*{Acknowledgments}
We gratefully acknowledge support from NVIDIA Corporation through an NVIDIA Academic Hardware Grant.
% \begin{acks}
% \hl{[COMPLETAR]}
% \end{acks}

% \balance
\bibliographystyle{ACM-Reference-Format}
\bibliography{references}


\begin{thebibliography}{11}

%%% the Web address.

\ifx \showCODEN    \undefined \def \showCODEN     #1{\unskip}     \fi
\ifx \showDOI      \undefined \def \showDOI       #1{#1}\fi
\ifx \showISBNx    \undefined \def \showISBNx     #1{\unskip}     \fi
\ifx \showISBNxiii \undefined \def \showISBNxiii  #1{\unskip}     \fi
\ifx \showISSN     \undefined \def \showISSN      #1{\unskip}     \fi
\ifx \showLCCN     \undefined \def \showLCCN      #1{\unskip}     \fi
\ifx \shownote     \undefined \def \shownote      #1{#1}          \fi
\ifx \showarticletitle \undefined \def \showarticletitle #1{#1}   \fi
\ifx \showURL      \undefined \def \showURL       {\relax}        \fi
% invisible to TeX
\providecommand\bibfield[2]{#2}
\providecommand\bibinfo[2]{#2}
\providecommand\natexlab[1]{#1}
\providecommand\showeprint[2][]{arXiv:#2}

\bibitem[Cheng et~al\mbox{.}(2016)]%
        {wdlrs16}
\bibfield{author}{\bibinfo{person}{Heng-Tze Cheng}, \bibinfo{person}{Levent
  Koc}, \bibinfo{person}{Jeremiah Harmsen}, \bibinfo{person}{Tal Shaked},
  \bibinfo{person}{Tushar Chandra}, \bibinfo{person}{Hrishi Aradhye},
  \bibinfo{person}{Glen Anderson}, \bibinfo{person}{Greg Corrado},
  \bibinfo{person}{Wei Chai}, \bibinfo{person}{Mustafa Ispir},
  \bibinfo{person}{Rohan Anil}, \bibinfo{person}{Zakaria Haque},
  \bibinfo{person}{Lichan Hong}, \bibinfo{person}{Vihan Jain},
  \bibinfo{person}{Xiaobing Liu}, {and} \bibinfo{person}{Hemal Shah}.}
  \bibinfo{year}{2016}\natexlab{}.
\newblock \showarticletitle{Wide \& Deep Learning for Recommender Systems}. In
  \bibinfo{booktitle}{\emph{Proceedings of the 1st Workshop on Deep Learning
  for Recommender Systems}} (Boston, MA, USA) \emph{(\bibinfo{series}{DLRS
  2016})}. \bibinfo{publisher}{Association for Computing Machinery},
  \bibinfo{address}{New York, NY, USA}, \bibinfo{pages}{7–10}.
\newblock
\showISBNx{9781450347952}
\urldef\tempurl%
\url{https://doi.org/10.1145/2988450.2988454}
\showDOI{\tempurl}


\bibitem[Cheng et~al\mbox{.}(2020)]%
        {afn_criteo}
\bibfield{author}{\bibinfo{person}{Weiyu Cheng}, \bibinfo{person}{Yanyan Shen},
  {and} \bibinfo{person}{Linpeng Huang}.} \bibinfo{year}{2020}\natexlab{}.
\newblock \showarticletitle{Adaptive Factorization Network: Learning
  Adaptive-Order Feature Interactions}. In
  \bibinfo{booktitle}{\emph{Proceedings of the AAAI Conference on Artificial
  Intelligence}}, Vol.~\bibinfo{volume}{34}. \bibinfo{pages}{3609--3616}.
\newblock
\urldef\tempurl%
\url{https://doi.org/10.1609/aaai.v34i04.5768}
\showDOI{\tempurl}


\bibitem[Gharibshah and Zhu(2021)]%
        {gharibshah2021user}
\bibfield{author}{\bibinfo{person}{Zhabiz Gharibshah} {and}
  \bibinfo{person}{Xingquan Zhu}.} \bibinfo{year}{2021}\natexlab{}.
\newblock \showarticletitle{User response prediction in online advertising}.
\newblock \bibinfo{journal}{\emph{aCM Computing Surveys (CSUR)}}
  \bibinfo{volume}{54}, \bibinfo{number}{3} (\bibinfo{year}{2021}),
  \bibinfo{pages}{1--43}.
\newblock


\bibitem[Guo et~al\mbox{.}(2017)]%
        {deepfm17}
\bibfield{author}{\bibinfo{person}{Huifeng Guo}, \bibinfo{person}{Ruiming
  TANG}, \bibinfo{person}{Yunming Ye}, \bibinfo{person}{Zhenguo Li}, {and}
  \bibinfo{person}{Xiuqiang He}.} \bibinfo{year}{2017}\natexlab{}.
\newblock \showarticletitle{DeepFM: A Factorization-Machine based Neural
  Network for CTR Prediction}. In \bibinfo{booktitle}{\emph{Proceedings of the
  Twenty-Sixth International Joint Conference on Artificial Intelligence,
  {IJCAI-17}}}. \bibinfo{pages}{1725--1731}.
\newblock
\urldef\tempurl%
\url{https://doi.org/10.24963/ijcai.2017/239}
\showDOI{\tempurl}


\bibitem[Harper and Konstan(2015)]%
        {movielens}
\bibfield{author}{\bibinfo{person}{F.~Maxwell Harper} {and}
  \bibinfo{person}{Joseph~A. Konstan}.} \bibinfo{year}{2015}\natexlab{}.
\newblock \showarticletitle{The MovieLens Datasets: History and Context}.
\newblock \bibinfo{journal}{\emph{ACM Trans. Interact. Intell. Syst.}}
  \bibinfo{volume}{5}, \bibinfo{number}{4}, Article \bibinfo{articleno}{19}
  (\bibinfo{date}{dec} \bibinfo{year}{2015}), \bibinfo{numpages}{19}~pages.
\newblock
\showISSN{2160-6455}
\urldef\tempurl%
\url{https://doi.org/10.1145/2827872}
\showDOI{\tempurl}


\bibitem[Hartford et~al\mbox{.}(2018)]%
        {dmias2018}
\bibfield{author}{\bibinfo{person}{Jason Hartford}, \bibinfo{person}{Devon
  Graham}, \bibinfo{person}{Kevin Leyton-Brown}, {and} \bibinfo{person}{Siamak
  Ravanbakhsh}.} \bibinfo{year}{2018}\natexlab{}.
\newblock \showarticletitle{Deep Models of Interactions Across Sets}. In
  \bibinfo{booktitle}{\emph{Proceedings of the 35th International Conference on
  Machine Learning}} \emph{(\bibinfo{series}{Proceedings of Machine Learning
  Research}, Vol.~\bibinfo{volume}{80})},
  \bibfield{editor}{\bibinfo{person}{Jennifer Dy} {and}
  \bibinfo{person}{Andreas Krause}} (Eds.). \bibinfo{publisher}{PMLR},
  \bibinfo{pages}{1909--1918}.
\newblock
\urldef\tempurl%
\url{https://proceedings.mlr.press/v80/hartford18a.html}
\showURL{%
\tempurl}


\bibitem[Mao et~al\mbox{.}(2023)]%
        {FinalMLP2023}
\bibfield{author}{\bibinfo{person}{Kelong Mao}, \bibinfo{person}{Jieming Zhu},
  \bibinfo{person}{Liangcai Su}, \bibinfo{person}{Guohao Cai},
  \bibinfo{person}{Yuru Li}, {and} \bibinfo{person}{Zhenhua Dong}.}
  \bibinfo{year}{2023}\natexlab{}.
\newblock \showarticletitle{FinalMLP: An Enhanced Two-Stream MLP Model for CTR
  Prediction}. In \bibinfo{booktitle}{\emph{Proceedings of the AAAI Conference
  on Artificial Intelligence}}, Vol.~\bibinfo{volume}{37}.
  \bibinfo{pages}{4552--4560}.
\newblock
\urldef\tempurl%
\url{https://doi.org/10.1609/aaai.v37i4.25577}
\showDOI{\tempurl}


\bibitem[Rendle(2010)]%
        {fm2010}
\bibfield{author}{\bibinfo{person}{Steffen Rendle}.}
  \bibinfo{year}{2010}\natexlab{}.
\newblock \showarticletitle{Factorization Machines}. In
  \bibinfo{booktitle}{\emph{2010 IEEE International Conference on Data
  Mining}}. \bibinfo{pages}{995--1000}.
\newblock
\urldef\tempurl%
\url{https://doi.org/10.1109/ICDM.2010.127}
\showDOI{\tempurl}


\bibitem[Weinsberg et~al\mbox{.}(2012)]%
        {gender_ml}
\bibfield{author}{\bibinfo{person}{Udi Weinsberg}, \bibinfo{person}{Smriti
  Bhagat}, \bibinfo{person}{Stratis Ioannidis}, {and} \bibinfo{person}{Nina
  Taft}.} \bibinfo{year}{2012}\natexlab{}.
\newblock \showarticletitle{BlurMe: Inferring and Obfuscating User Gender Based
  on Ratings}. In \bibinfo{booktitle}{\emph{Proceedings of the Sixth ACM
  Conference on Recommender Systems}} (Dublin, Ireland)
  \emph{(\bibinfo{series}{RecSys '12})}. \bibinfo{publisher}{Association for
  Computing Machinery}, \bibinfo{address}{New York, NY, USA},
  \bibinfo{pages}{195–202}.
\newblock
\showISBNx{9781450312707}
\urldef\tempurl%
\url{https://doi.org/10.1145/2365952.2365989}
\showDOI{\tempurl}


\bibitem[{Zamanzadeh Darban} and Valipour(2022)]%
        {GHRS2022}
\bibfield{author}{\bibinfo{person}{Zahra {Zamanzadeh Darban}} {and}
  \bibinfo{person}{Mohammad~Hadi Valipour}.} \bibinfo{year}{2022}\natexlab{}.
\newblock \showarticletitle{GHRS: Graph-based hybrid recommendation system with
  application to movie recommendation}.
\newblock \bibinfo{journal}{\emph{Expert Systems with Applications}}
  \bibinfo{volume}{200} (\bibinfo{year}{2022}), \bibinfo{pages}{116850}.
\newblock
\showISSN{0957-4174}
\urldef\tempurl%
\url{https://doi.org/10.1016/j.eswa.2022.116850}
\showDOI{\tempurl}


\bibitem[Zhu et~al\mbox{.}(2022)]%
        {bars_ds}
\bibfield{author}{\bibinfo{person}{Jieming Zhu}, \bibinfo{person}{Quanyu Dai},
  \bibinfo{person}{Liangcai Su}, \bibinfo{person}{Rong Ma},
  \bibinfo{person}{Jinyang Liu}, \bibinfo{person}{Guohao Cai},
  \bibinfo{person}{Xi Xiao}, {and} \bibinfo{person}{Rui Zhang}.}
  \bibinfo{year}{2022}\natexlab{}.
\newblock \showarticletitle{BARS: Towards Open Benchmarking for Recommender
  Systems}. In \bibinfo{booktitle}{\emph{Proceedings of the 45th International
  ACM SIGIR Conference on Research and Development in Information Retrieval}}
  (Madrid, Spain) \emph{(\bibinfo{series}{SIGIR '22})}.
  \bibinfo{publisher}{Association for Computing Machinery},
  \bibinfo{address}{New York, NY, USA}, \bibinfo{pages}{2912–2923}.
\newblock
\showISBNx{9781450387323}
\urldef\tempurl%
\url{https://doi.org/10.1145/3477495.3531723}
\showDOI{\tempurl}


\end{thebibliography}

\end{document}